\documentclass[10pt]{iopart}
\usepackage{epsfig}
\usepackage{amstext}
\usepackage{amssymb}
\usepackage{amsfonts}
\usepackage{graphicx}
\usepackage{vmargin}
\usepackage{amscd}
\usepackage{bbold}
\usepackage{mathrsfs}
\usepackage{fancyheadings}
\usepackage{pifont}
\usepackage{pst-node}

\newcommand{\be}{\begin{equation}}
\newcommand{\ee}{\end{equation}}
\begin{document}
\title{Effects of In-Medium NN Cross Sections on Particle Production}

\author{V Prassa$^{1}$, T Gaitanos$^{2}$, H H Wolter$^{2}$, 
G Lalazissis$^{1}$, M Di Toro$^{3}$}

\address{$^{1}$ Department of Theoretical Physics, 
Aristotle University of Thessaloniki, Thessaloniki Gr-54124,Greece}
\address{$^{2}$ Dept. f\"{u}r Physik, Universit\"{a}t M\"{u}nchen, 
Am Coulombwall 1, D-85748 Garching, Germany}
\address{$^{3}$ Laboratori Nazionali del Sud INFN, I-95123 Catania, Italy}
%\author{\begin{quote}
\begin{abstract}
The in-medium modification of nucleon-nucleon (NN) cross sections is 
investigated by means of particle production in heavy ion collisions (HIC)
at intermediate energies. In particular, the density dependence of the 
{\it inelastic} cross sections considerably affects the pion and kaon 
yields and their rapidity distributions. However, the 
$(\pi^{-}/\pi^{+})$- and $(K^{+}/K^{0})$-ratios depend only moderately 
on the in-medium behavior of the inelastic cross sections. It turns out 
that particle ratios seem to be robust observables in determining the 
nuclear equation of state (EoS) and, particularly, its isovector sector.
\end{abstract}
%\end{quote}}
\date{\today}
%\maketitle
%\setpapersize{A4}
      %    Width of rule between columns.
%%%%%%%%%%%%%%%%%%%%%%%%%%%%%%%%%%%%%%%%%%%%%%%%%%%%%%%%%%%%%%%%%%%
%%%%%%%%%%%%%%%%%%%%%%%%%%%%%%%%%%%%%%%%%%%%%%%%%%%%%%%%%%%%%%%%%%%      
%%%%%%%%%%%%%%%%%%%%%%%%%%%%%%%%%%%%%%%%%%%%%%%%%%%%%%%%%%%%%%%%%%%
%	 		BEGIN OF TEXT				  %
%%%%%%%%%%%%%%%%%%%%%%%%%%%%%%%%%%%%%%%%%%%%%%%%%%%%%%%%%%%%%%%%%%%
%%%%%%%%%%%%%%%%%%%%%%%%%%%%%%%%%%%%%%%%%%%%%%%%%%%%%%%%%%%%%%%%%%%
%%%%%%%%%%%%%%%%%%%%%%%%%%%%%%%%%%%%%%%%%%%%%%%%%%%%%%%%%%%%%%%%%%%   

%%%%%%%%%%%%%%%%%%%%%%%%%%%%%%%%%%%%%%%%%%%%%%%%%%%%%%%%%%%%%%%%%%%   
\section{Introduction}
%%%%%%%%%%%%%%%%%%%%%%%%%%%%%%%%%%%%%%%%%%%%%%%%%%%%%%%%%%%%%%%%%%%
The knowledge of the properties of highly compressed and heated hadronic 
matter is an important issue for the understanding of astrophysics such as 
the physical mechanism of supernovae explosions and the physics of neutron 
stars \cite{NS1,NS2}. HIC provide the unique opportunity to explore highly excited 
hadronic matter, i.e. the high density behavior of the nuclear EoS, under 
controlled conditions (high baryon energy densities and temperatures) 
in the laboratory \cite{dani}. 

Important observables have been the nucleon collective dynamics 
\cite{dani,ritter} and the dynamics of produced particles such as pions and 
kaons \cite{fuchs}. However, the reaction dynamics is a rather 
complex process which involves the nuclear mean field (EoS) and binary 
$2$-body collisions. In the presence of the nuclear medium  
the treatment of binary collisions represents a non-trivial topic. 
The NN cross sections for elastic and inelastic processes, which are the 
crucial physical parameters here, are experimentally accessible only for the 
free space and not for $2$-body scattering at finite baryon density. Recent 
microscopic studies, based on the $T$-matrix approach, have shown a strong 
decrease of the elastic NN cross section \cite{fuchs2} in the presence of a 
hadronic medium. These in-medium effects of the elastic NN cross 
section considerably influence the hadronic reaction dynamics \cite{gaitcross}. 
Obviously the question arises whether similar in-medium effects of the {\it 
inelastic} NN cross sections may affect the reaction dynamics and, in 
particular, the production of particles (pions and kaons). 

Since microscopic results are not available, 
we discuss here in a simple phenomenological way possible density modifications 
of the inelastic NN cross sections and their influences on particle 
multiplicities, rapidity distributions and ratios. 
We find a strong dependence of the yields and rapidity distributions 
on the in-medium modifications of the inelastic cross sections, but on the 
other hand, this effect is only moderate for particle ratios such as 
$(\pi^{-}/\pi^{+})$, and almost vanishes for $(K^{+}/K^{0})$. 
Therefore such ratios turn out 
to be robust observables in determining the nuclear EoS and, particularly, 
the isovector channel of the nuclear mean field \cite{ferini,stoecker,bao}.

%%%%%%%%%%%%%%%%%%%%%%%%%%%%%%%%%%%%%%%%%%%%%%%%%%%%%%%%%%%%%%%%%%%
\section{The transport equation}
%%%%%%%%%%%%%%%%%%%%%%%%%%%%%%%%%%%%%%%%%%%%%%%%%%%%%%%%%%%%%%%%%%%

In this chapter we briefly discuss the transport equation by concentrating 
on the treatment of the cross sections, which are the important parameters 
of the collision integral. 

The theoretical description of HIC is based on the kinetic theory of 
statistical mechanics, i.e. the Boltzmann Equation \cite{kada}. The 
relativistic semi-classical analogon of this equation is the 
Relativistic Boltzmann-Uehling-Uhlenbeck (RBUU) equation \cite{giessen} 
%%%%%%%%%%%%%%%
\begin{eqnarray}
& & \left[ 
k^{*\mu} \partial_{\mu}^{x} + \left( k^{*}_{\nu} F^{\mu\nu} 
+ M^{*} \partial_{x}^{\mu} M^{*}  \right) 
\partial_{\mu}^{k^{*}} 
\right] f(x,k^{*}) = \frac{1}{2(2\pi)^9} \nonumber\\
& & \times \int \frac{d^3 k_{2}}{E^{*}_{{\bf k}_{2}}} 
             \frac{d^3 k_{3}}{E^{*}_{{\bf k}_{3}}}
             \frac{d^3 k_{4}}{E^{*}_{{\bf k}_{4}}} W(kk_2|k_3 k_4)   
 \left[ f_3 f_4 \tilde{f}\tilde{f}_2 -f f_2 \tilde{f}_3\tilde{f}_4 
\right]
\label{rbuu} 
\end{eqnarray}
%%%%%%%%%%%%%%%
where $f(x,k^{*})$ is the single particle distribution function. 
In the collision term the short-hand notation $f_i \equiv f(x,k^{*}_i)$ 
for the particle and $\tilde{f}_i \equiv (1-f(x,k^{*}_i))$ 
and the hole-distribution is used. 
The collision integral exhibits explicitly the final state Pauli-blocking while 
the in-medium scattering amplitude includes the  Pauli-blocking of intermediate 
states. %%%%%%%%%%%%%%%%%%%%%%%%%%%%%%%%%%%%%%%%%%%%%%%%%%%%%%%%%%%%%%%%%%%
\begin{figure}[t]
\unitlength1cm
\begin{picture}(8.,7.3)
\put(3.0,0.3){\makebox{\epsfig{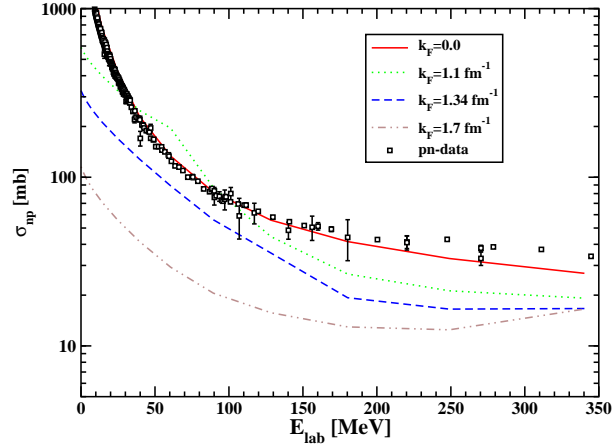}}}
\end{picture}
\caption{Elastic in-medium neutron-proton cross sections 
at various Fermi momenta 
$k_{F}$ as function of the laboratory energy $E_{lab}$. The free cross
section ($k_{F}=0$) is compared to the experimental total $np$ cross
section \protect\cite{fuchs2}.
}
\label{fig1}
\end{figure}
%%%%%%%%%%%%%%%%%%%%%%%%%%%%%%%%%%%%%%%%%%%%%%%%%%%%%%%%%%%%%%%%%%%
The dynamics of the lhs of eq.(\ref{rbuu}), the drift term, is 
determined by the mean field. Here the 
attractive scalar field $\Sigma_S$ enters via the effective mass 
$M^{*}=M-\Sigma_{s}$ 
and the repulsive vector field $\Sigma_\mu$ via 
kinetic momenta $k^{*}_{\mu}=k_{\mu}-\Sigma_{\mu}$ and via the field tensor 
$F^{\mu\nu} = \partial^\mu \Sigma^\nu -\partial^\nu \Sigma^\mu$. 
The in-medium cross sections enter into the collision integral 
via the transition amplitude  
%%%%%%%%%%%%%%%
\begin{equation}
W = (2\pi)^4 \delta^4 \left(k + k_{2} -k_{3} - k_{4} \right) 
(M^*)^4 |T|^2~~
\label{trans}
\end{equation}
%%%%%%%%%%%%%%%
with $T$ the in-medium scattering matrix element. 

In the kinetic equation (\ref{rbuu}) one should use both physical 
quantities, the mean field (EoS) and the collision integral (cross sections) 
according to the same underlying effective 
two-body interaction in the medium,
i.e. the in-medium T-matrix;  $\Sigma \sim \Re T\rho,~~
\sigma\sim \Im T$, respectively $W \sim | T|^2$. 
However, in most practical applications phenomenological mean fields 
and cross sections have been used. In these models adjusting the 
known bulk properties 
of nuclear matter around the saturation point one tries to 
constrain the models for supra-normal densities 
with the help of heavy ion reactions 
\cite{dani00,larionov00}. Medium modifications of the NN cross section 
are usually not taken into account which works, in comparison to 
experimental data, 
astonishingly well \cite{dani00,larionov00,gait01,gaitacnm}. 
However, in particular kinematics regimes a 
sensitivity of dynamical observables such as collective flow and 
stopping \cite{gaitcross,gale} or transverse energy transfer 
\cite{dancross} to the elastic NN cross section has been observed. 

%%%%%%%%%%%%%%%%%%%%%%%%%%%%%%%%%%%%%%%%%%%%%%%%%%%%%%%%%%%%%%%%%%%
\begin{figure}[t]
\unitlength1cm
\begin{picture}(8.,7.3)
\put(3.0,0.3){\makebox{\epsfig{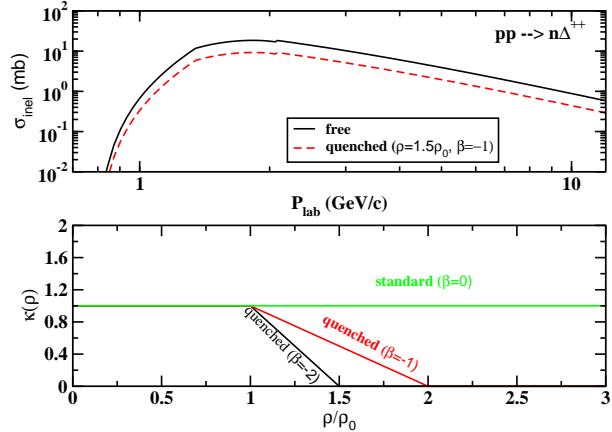}}}
\end{picture}
\caption{Top: inelastic cross section in free (solid) space and 
in the medium ($\rho=1.5 \rho_{0}$) using $\beta=-1$. 
Bottom: Density dependence of the quenching function 
$\kappa(\rho,\beta)$ for different $\beta$-values.
}
\label{fig2}
\end{figure}
%%%%%%%%%%%%%%%%%%%%%%%%%%%%%%%%%%%%%%%%%%%%%%%%%%%%%%%%%%%%%%%%%%%
Fig. \ref{fig1} shows the energy dependence of the in-medium 
neutron-proton $(np)$ cross section at 
Fermi momenta $k_F = 0.0, 1.1,1.34,1.7 fm^{-1}$, 
corresponding to $\rho \sim 0,0.5,1,2\rho_0$ ($\rho_0=0.16 fm^{-3}$ 
is the nuclear matter saturation density) as found in relativistic 
Dirac-Brueckner (DB) calculations \cite{fuchs2}. 
The presence of the medium leads to a substantial 
suppression of the cross section which is most pronounced at 
low laboratory energy $E_{\rm lab}$ and high densities where 
the Pauli-blocking of intermediate states is 
most efficient. 
At larger $E_{\rm lab}$ asymptotic values of 
15-20 mb are reached. However, not only the total cross section 
but also the angular distributions are affected by the presence of the 
medium. The initially strongly forward-backward 
peaked  $np$ cross sections become much more isotropic at finite densities 
\cite{fuchs2} which is mainly do to the Pauli suppression of 
soft modes ($\pi$-exchange) and correspondingly of higher partial waves in 
the T-matrix \cite{fuchs2}.

Obviously one expects similar in-medium effects for the 
inelastic NN cross sections mainly due to Pauli-blocking of intermediate 
scattering states and in-medium modified matrix elements. Such microscopic 
studies for inelastic processes are very rare or still in development 
\cite{lari1,lari2}. However, to explore the sensitivity we use here a rather simple 
parametrization which assumes a reduction of the inelastic NN cross section with 
increasing baryon density, in line with that of Fig. \ref{fig1} for the elastic 
one which has previously been used in Ref. \cite{lari1}. 
This can be achieved by assuming a factorization of the effective 
matrix element of the form 
%%%%%%%%%%%%%%%
\begin{equation}
\overline{|M_{eff}|^{2}} = 
\kappa(\rho,\beta)\overline{|M_{vac}|^{2}}
\label{meff}
\end{equation}
%%%%%%%%%%%%%%%
with $M_{vac}$ the vacuum matrix element (taken from experimental free 
scattering data) and $\kappa(\rho,\beta)$ a density dependent function 
depending on a quenching parameter $\beta$. 
A very simple parametrization of the function $\kappa(\rho)$ is 
shown in Fig. \ref{fig2} (bottom panel) for different values of the parameter 
$\beta$. The effect on the inelastic NN cross section is shown 
on the top of Fig. \ref{fig2} for the choice $\beta=-1$. A significant reduction 
of the effective inelastic NN cross section (by a factor of 2) is observed with 
respect to that of the free case at a given baryon density $\rho=1.5\rho_{0}$. 

%%%%%%%%%%%%%%%%%%%%%%%%%%%%%%%%%%%%%%%%%%%%%%%%%%%%%%%%%%%%%%%%%%%
\section{Results}
%%%%%%%%%%%%%%%%%%%%%%%%%%%%%%%%%%%%%%%%%%%%%%%%%%%%%%%%%%%%%%%%%%%

We have applied the parametrization (\ref{meff}) in the collision integral 
of the transport equation (\ref{rbuu}) and analyzed the transport 
calculations in terms of particle production. At energies below 
1.6 AGeV pions and kaons are produced, where the second ones do not 
significantly affect the reaction dynamics due to their small production 
cross sections. In particular, the kaons are treated perturbatively 
and can be produced from $BB \longrightarrow BYK$ and 
$\pi B \longrightarrow YK$. Here $B$ stands for a nucleon or 
$\Delta$ resonance and $Y$ for a hyperon. Pions are created via the 
decay of the $\Delta(1232)$ resonance. For more details we 
refer to \cite{ferini}. For the nuclear mean field the $NL2$ parametrization 
of the non-linear Walecka model \cite{mosel} is adopted here with a 
compression modulus of $200$ MeV and an effective Dirac mass of 
$m^{*}=0.82~M$ ($M$ is the bare nucleon mass). The momentum dependence 
enters via the relativistic treatment through the vector components of 
the baryon self energy. In order to keep the discussion transparent, we 
do not apply here any isovector components of the baryon self energy, 
as it has been previously done in Ref. \cite{ferini}. This implies no 
effective mass splitting between protons and neutrons, and on the other 
hand, between the different isospin states of the $\Delta$ resonance 
($\Delta^{-,0},~\Delta^{+,++}$) and of the hyperons 
($\Lambda,~\Sigma^{\pm,0}$). Furthermore, pions are propagated in a 
coulomb field and kaons do not experience any potential. 

%%%%%%%%%%%%%%%%%%%%%%%%%%%%%%%%%%%%%%%%%%%%%%%%%%%%%%%%%%%%%%%%%%%
\begin{figure}[t]
\unitlength1cm
\begin{picture}(8.,7.)
\put(3.0,0.3){\makebox{\epsfig{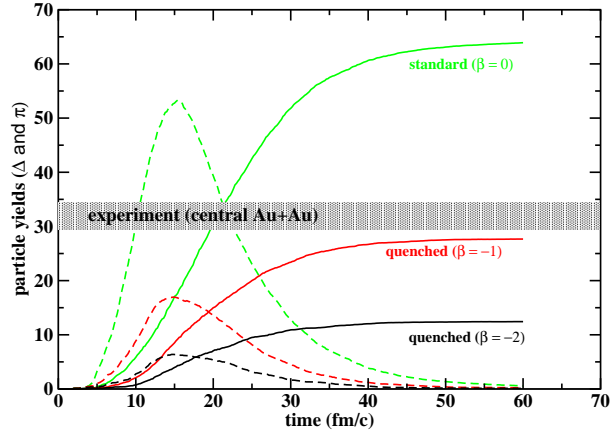}}}
\end{picture}
\caption{Time evolution (in units of $fm/c$) 
of the multiplicity of $\Delta$-resonances (dashed lines) and pions 
(solid lines) for different $\beta$-values as indicated. The gray band 
represents the range of experimental 
data for central Au+Au reactions at 1 AGeV incident energy \cite{pelte}.}
\label{fig3}
\end{figure}
%%%%%%%%%%%%%%%%%%%%%%%%%%%%%%%%%%%%%%%%%%%%%%%%%%%%%%%%%%%%%%%%%%%
We start the discussion with the time evolution of the multiplicities 
of the $\Delta$ resonances and of the pions, as can be seen in 
Fig. \ref{fig3}. 
The time evolution of the multiplicity of produced $\Delta$-resonances is shown with 
their maximum around 15 fm/c which corresponds to the time of 
maximum compression. Due to the finite lifetime these resonances decay into 
pions (and nucleons) according $\Delta \longrightarrow \pi N$ (some of these 
pions are re-absorbed in the inverse process, 
i.e. $\pi N \longrightarrow \Delta$). This mechanism continues until all 
resonances have decayed leading to a final constant pion yield for times $t \geq 50$ 
fm/c (the so-called freeze-out time). After the freeze-out the pions can be 
measured experimentally. The experimental pion multiplicity is 
schematically shown in Fig. \ref{fig2} by the gray band for central Au+Au 
collisions \cite{pelte}. 
We observe an essential reduction of the pion multiplicity using the 
effective inelastic NN cross sections, in line with the previous Fig. 
\ref{fig2}. As an important result the calculations with the effective inelastic 
NN cross section describe the experimental data reasonably well for 
a quenching parameter of about $\beta=-1$. However, for a more realistic comparison 
one should use more microscopic calculations of the inelastic cross section, 
as it has been done in Ref. \cite{lari2}. Such a progress is under study. 

\begin{figure}[t]
\unitlength1cm
\begin{picture}(8.,7.)
\put(3.0,0.3){\makebox{\epsfig{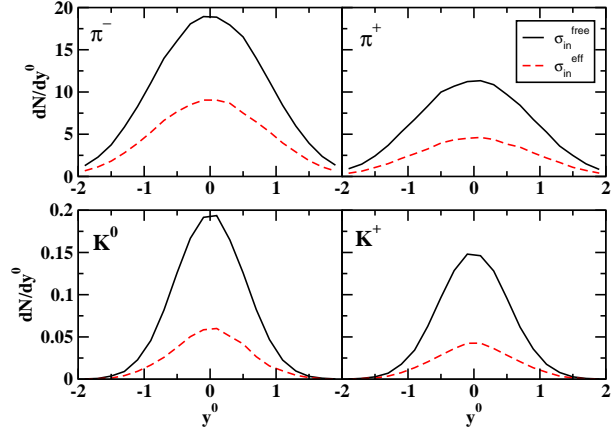}}}
\end{picture}
\caption{Rapidity distributions of $\pi^{-}$ (top-left), 
$\pi^{+}$ (top-right), $K^{0}$ (bottom-left) and 
$K^{+}$ (bottom-right) for central ($b=1$ fm) Au+Au collisions at 
1.48 AGeV incident energy.Results of transport calculations using the 
free (Solid lines) and in-medium (dashed lines) inelastic cross sections with 
$\beta=-1$ are shown.}
\label{fig4}
\end{figure}

The strong in-medium dependence of the inelastic cross sections is shown also 
in the rapidity distributions of pions and kaons in Fig. \ref{fig4}. 
%%%%%%%%%%%%%%%%%%%%%%%%%%%%%%%%%%%%%%%%%%%%%%%%%%%%%%%%%%%%%%%%%%%
%%%%%%%%%%%%%%%%%%%%%%%%%%%%%%%%%%%%%%%%%%%%%%%%%%%%%%%%%%%%%%%%%%%
The rapidity distribution of kaons is even more stronger affected by the 
density behavior of the inelastic cross section with respect to 
that of pions. This is due to the fact that the leading channels for 
kaon production are $N\Delta \longrightarrow BYK$ and pionic ones, which 
are both reduced when the in-medium dependent $\sigma_{inel}$ are used.

The question arises whether particle ratios are influenced by 
in-medium effects of the inelastic cross section. An answer on this 
question is of major importance, particularly for kaons and less for 
pions, since particle rations have been widely used in determining 
the nuclear EoS at supra-normal densities. relative ratios of kaons 
between different colliding systems have been used in determining the 
isoscalar sector of the nuclear EoS, see Refs. \cite{fuchs}. More 
recently, the $(\pi^{-}/\pi^{+})$- and $(K^{+}/K^{0})$-ratios have 
been used in exploring the high density behavior of the symmetry 
energy, i.e. the isovector part of the nuclear mean field \cite{bao,stoecker,ferini}. 

\begin{table}[hbt]
\centering
\begin{tabular}{|c||c|c|} \hline
particle ratio & free $\sigma_{inel}$ & in-medium $\sigma_{inel} (\beta=-1)$ \\ \hline\hline
$(\pi^{-}/\pi^{+})$ & 1.63 ($\pm$ 0.11) & 1.92 ($\pm$ 0.14) \\ \hline
$(K^{+}/K^{0})$ & 0.77 ($\pm$ 0.05)  & 0.74 ($\pm$ 0.05) \\ \hline
\end{tabular}
\caption{In-medium dependence of the $(\pi^{-}/\pi^{+})$- and $(K^{+}/K^{0})$-ratios 
for central ($b=1$ fm) Au+Au collisions at 1.48 AGeV incident energy. }
\label{table1}
\end{table}
Table \ref{table1} shows the $(\pi^{-}/\pi^{+})$- and 
$(K^{+}/K^{0})$-ratio for central ($b=1$ fm) Au+Au collisions at 
1.48 AGeV incident energy using both, the free and in-medium 
inelastic cross sections. The pionic ratio moderately 
depends on the density dependence of the inelastic cross section, 
but the $(K^{+}/K^{0})$-ratio is not particularly affected by the 
in-medium effects. A possible reason for the different behavior between 
the pionic and kaonic ratios may be the fast pre-equilibrium emission of 
the kaons. In particular, kaons are created very early during the formation 
of the high density phase and are emitted from the compression region 
without undergoing any interaction with the hadronic environment. Therefore 
one obviously expects a direct relation between the high density effects of the inelastic 
cross sections and the $(K^{+}/K^{0})$-ratio. Pions, 
on the other hand, are strongly interacting with the medium via secondary 
re-absorption processes and thus are emitted from different stages of a 
collision which may influence the final $(\pi^{-}/\pi^{+})$-ratio \cite{gait04,bao}. 
It turns out that the strangeness ratio presents a robust observable to investigate 
the isovector character of the nuclear matter EoS. 

%%%%%%%%%%%%%%%%%%%%%%%%%%%%%%%%%%%%%%%%%%%%%%%%%%%%%%%%%%%%%%%%%%%
\section{Conclusions}
%%%%%%%%%%%%%%%%%%%%%%%%%%%%%%%%%%%%%%%%%%%%%%%%%%%%%%%%%%%%%%%%%%%

We have investigated the role of the density dependence of the inelastic cross 
section on particle production in intermediate energy heavy ion collisions 
within a covariant transport equation of a Boltzmann type. Since microscopic 
studies on the in-medium behavior of inelastic cross sections are still rare, 
we have used here a simple phenomenological in-medium dependence of the 
inelastic cross sections. We have applied the transport equation to Au+Au 
collisions at intermediate relativistic energies below the kaon threshold energy. 

Our studies have shown a strong sensitivity of the particle multiplicities and 
rapidity distributions of pions and kaons. In particular, a reduction by a 
factor of $2$ for pions has been seen when the in-medium effects in the inelastic 
cross section are accounted for. Consequently, the kaon ($K^{0,+}$) yields decrease 
more than 50 \%. This is due to the asumption of a reduction of the inelastic 
cross section at high densities. As an interesting finding, the multiplicities 
of $K^{0}$ and $K^{+}$ are influenced in such a way that their ratio is independent 
on the density dependence of the inelastic cross sections. This may be due to the 
long mean free path of the $K^{0,+}$, whereas the pionic ratio, due to their 
strong secondary interaction processes with the hadronic environment, have shown 
a moderate dependence on the density behavior of the inelastic cross sections. 

Certainly more systematic studies are necessary to investigate better the mechanism 
which leads to a moderate relation between the in-medium dependence of the inelastic 
cross sections and the pionic ratio. A comparison with more experimental data would 
be helpful in determining more precisely the density dependence of the inelastic 
cross sections, which will be an object of future studies. On this level of investigations 
we conclude that the $(K^{+}/K^{0})$-ratio represents a robust observable in determining 
the nuclear matter EoS at supra-normal densities. \\

{\it Acknowledgment} This work is supported by BMBF, grand 06LM189. 
One of the authors (V.P.) would like to thank 
H.H. Wolter and M. Di Toro for the warm hospitality during her short 
stays at their institutes.

\end{document}